\documentstyle[12pt,epsfig]{article}

 \def\four{{\textstyle{1\over 4}} }
 \def\ha{{\textstyle{1\over 2}} }
 \def\p{\partial}
 
 \def\a{\alpha}
\def\b{\beta }
\def\g{\gamma }
\def\s{\sigma }
\def\vp{\varphi }

\def\E{ {\cal E}}

\def\non{\nonumber }
\def\beq{\begin{equation} }
\def\eeq{\end{equation} }
\def\beqa{\begin{eqnarray}}
\def\eeqa{\end{eqnarray}}

\begin{document}

\begin{titlepage}
\hfill
\vbox{
    \halign{#\hfil         \cr
           SISSA 7/2002/EP \cr
           hep-th/0201269   \cr
           } 
      }  
\vspace*{15mm}
\begin{center}
{\Large {\bf Quantum entanglement of charges in\\
            bound states with finite-size dyons} }\\

\vspace*{12mm}

{Roberto Iengo{}~$^a$ and Jorge G. Russo{}~$^b$}\\

\vspace*{0.6cm}

{\it ${}^a$~International School for Advanced Studies (SISSA),
via Beirut 2-4, I-34013 Trieste, Italy\\
INFN, Sezione di Trieste.}\\
{\tt iengo@sissa.it} \\

\medskip
{\it $^b$ Departamento de F\'\i sica, Universidad de Buenos Aires\\
Ciudad Universitaria, 1428 Buenos Aires, and Conicet.}\\
 {\tt russo@df.uba.ar} \\

\vspace*{0.5cm}
\end{center}

\begin{abstract}

  We show that the presence of finite-size monopoles can  lead to a number of interesting
physical processes involving quantum entanglement of charges.
Taking as a model the classical solution of the $N=2$  $SU(2)$
Yang-Mills theory, we study interaction between dyons and scalar
particles in the adjoint and fundamental representation. We find
that there are bound states of scalars and  dyons, which,
remarkably,  are always an entangled configuration  of the form
$|\psi \rangle =|{\rm dyon}_+\rangle |{\rm scalar}_-\rangle \pm
|{\rm dyon}_-\rangle |{\rm scalar}_+\rangle $. We determine the
energy levels and the wave functions and also discuss their
stability.

\end{abstract}
\vskip 0.5cm

January 2002
\end{titlepage}

\newpage

\section{Introduction}
The presence of a 't Hooft-Polyakov monopole
\cite{HP,Poly} gives rise to
remarkable effects such as baryon number violation \cite{caru,ruba}.
It is also believed that there must exist processes where a charged
particle transfers  charge to the monopole, converting it into a dyon
(see e.g.  \cite{blaer,bala,yoneya,kazama}), or more generally, that there are charge transfer processes in the interaction of particles with dyons.
These effects are believed to be quite generic,
peculiar to the nature of the monopole field,
but otherwise model
independent. In particular, they are present in  supersymmetric models.

Here we will consider  specifically
the Prasad-Sommerfield dyon solution
of $N=2$ supersymmetric $SU(2)$ Yang-Mills theory \cite{PS}
(for reviews see  \cite{olive,harvey}) and its interaction with scalar
particles of a $SU(2)$ multiplet in the adjoint and in the fundamental
representation.

In general, the  equations of the scalar field in the background of
the dyon couple
the different members of the $SU(2)$ multiplet.
By defining the electric charge in terms of the
asymptotic states where the Higgs field is constant, far from the dyon core,
this fact is seen as a nonconservation of the charge of the scalar field.
Since it is possible to define the {\it total} electric charge by the
flux at infinity of the total electric field
(that is, including the electric field of the dyon solution) and show that
it is conserved, this implies that in the process there is a transfer
of charge from the particle to the dyon core.
While it is difficult to incorporate exactly the back reaction of
the scalar field on the dyon solution, it is possible  to take into account
the main effect of the total charge conservation by representing the
charge degrees of freedom of the dyon by means of a quantum rotator formalism
(\cite{coleman,harvey}, see in particular ref.\cite{bala}).
The outcome is that in general
the quantum states of the particle-dyon system are of the form of
quantum entanglement, i.e. a linear superposition of
particle-dyon states in which the total charge is fixed, but the particle and
the dyon appear with various charge assignements.

It is interesting to compare with the string theory description of the $N=2$ monopole in terms of D branes.
 The $N=2$ $SU(2)$ monopole can be geometrically
described as a  tube (representing a D string) connecting
two parallel D3 branes \cite{hashimoto}.
The vacuum
expectation value of the Higgs field is proportional to the
distance between the D3 branes. This separation breaks the
$SU(2)$ gauge group  to $U(1)$.
Consider a small open string representing a neutral particle
on one of the two D3-branes: one of the string end points, say the
negatively charged one, can fall
into the monopole and pass through it to the other brane, whereas
the other endpoint may remain on the first brane. An observer on the first brane would see a positively charged particle and a negative charge dyon,
due to the flux lines of the electric field dragged by the particle which falled into the monopole. The string would go from the positive particle to
the dyon
(providing a physical realization of the mathematical string
introduced in ref.\cite{kazama} for preserving gauge invariance)
and pass through the tube up to the negative end point on the other brane.
Clearly, it is equally probable that the positively charged endpoint of the original open string falls into the monopole, which would lead to a system of a positive charge dyon and a negative charge particle on the first brane.
Of course this geometrical picture is unable to take into account
the quantum effect of entanglement, whereby the resulting state
can be  a superposition of the  two possible realities: a
dyon (+) and a particle ($-$), or a dyon ($-$) and a particle (+).

\bigskip

In this paper we consider
the possible bound states of a dyon and a
charged scalar particle. We first perform a general harmonic analysis in the case
of the finite-size monopole
and, after diagonalizing  the angular momentum, we get
coupled radial equations, where the off-diagonal terms
--~representing
the coupling of different charges~-- vanish exponentially outside the dyon core.
As a result, we find that the possible bound states are always
of the entangled form.
In other words, a bound state just of the form, say,
$|{\rm dyon}_-\rangle |{\rm scalar}_+\rangle$
(a plus charged scalar particle and a minus charged dyon)
can never occur.
Rather, the possible bound states are
a linear combination of the above state with the state of opposite charge, $|{\rm dyon}_+\rangle |{\rm scalar}_-\rangle$.
At large distances, the system is similar to a hydrogen atom, with
a particle orbiting far away from the core.
We will find that the energy spectrum and wave functions can indeed be approximated  by
the same formulas of the hydrogen atom, with a proper definition of the effective charge and effective
angular momentum.  The most relevant effect of the core is thus
the production of quantum entanglement.

Bound states of particles with
dyons have been studied in the past  (see  e.g.
\cite{tang,ajith}), mainly for fermions and in the point-like core
limit, except for the special zero energy state found by Jackiw
and Rebbi \cite{jackiw}. However, the peculiar fact that bound
states to dyons are always quantum entangled with the system of
opposite charges does not emerge in this limit. In ref.
\cite{raven1, raven2} a more refined analysis was performed, by
keeping into account the core interior in some stepwise
approximation, but focusing on the case of the monopole or for a
fixed dyon, and mainly for investigations on the Jackiw-Rebbi
phenomenon; dynamical effects of charge transfer and related
collective dyon modes were not taken into account.

In order to observe the quantum entanglement effect, the key points are:
a) the interactions in the monopole interior (in particular,
the effect would not appear in a large distance approximation
neglecting what happens inside the core); b) the account of the dyon degrees of freedom
and of the collective mode
which is responsible for charge conservation.
Here we report  on this phenomenon, not only by controlled analytic
approximations but also providing the exact numerical solutions to the bound
state problem.

Our computation is done for the case of globally neutral systems of
scalars and dyons, for scalars in the adjoint and fundamental representation of $SU(2)$,
for the case of scalars belonging to the $N=2$ supermultiplet and also more in general.
We also discuss briefly the stability of the bound states against
 perturbations not included in our computation, in particular,
the radiation of e.m. quanta.
A summary of the results is reported in the last Section 5.

\section{ Dyons in  $N=2$ $SU(2)$ super Yang-Mills}

\subsection{Monopole and dyon solutions}

 Let $\sigma^a ,\  a=1,2,3$
be the Pauli matrices, satisfying $[\sigma^a,\sigma^b
]=2i\epsilon^{abc}\sigma^c$.
The background describing the
dyon of $N=2$ $SU(2)$ super Yang-Mills
theory is given by \cite{PS,olive}
\beq
A^i=A^i_b\sigma ^b = - \sigma_b\epsilon_{bij}{r^j\over e r^2}\ f(x)  \ ,\
\ \ \
 \ \Phi_b\sigma^b ={\vec r.\vec\sigma \over r} a G(x)\ ,
\label{due}
\eeq
\beq
A^0={\vec r.\vec\s\over 2r} a G(x) \sin\Theta \ ,\ \ \ x\equiv ear \cos\Theta\ ,
\label{reso}
\eeq
$$
f(x)=-\ha (1-K(x))\ ,\ \ K(x)={x\over {\rm sinh}(x)}\ ,\ \ \ \
G(x)={\rm coth}x-{1\over x}\ ,
$$
where $a$ represents the expectation value of a scalar field, and $(ea\cos\Theta )^{-1}$  determines
the dyon size.
The behavior at large $x$ and small $x$ is
\beq
  K(x)=O(e^{-x})\ ,\ \ \
G(x)=1-{1\over x}+O(e^{-x})\ ,\ \ \ \ x\gg 1\ ,
\label{cou}
\eeq
\beq
  K(x)=1-{x^2\over 6}+O(x^4)   \ ,\ \ \
G(x)={x\over 3}+O(x^3)\ ,        \ \ \ x\ll 1\ .
\label{reg}
\eeq
The dyon charge and mass are given by
$$
q=g \tan\Theta =-{4\pi\over e} \tan\Theta\ ,\ \   \ q=n  e\ ,
$$
\beq
M=a\sqrt{g^2+n^2 e^2}  =\bigg|{ag\over \cos\Theta }\bigg|\ .
\label{dyma}
\eeq
 For  $e^2 \ll 1$, this becomes
\beq
M \cong |ag|(1+\ha \a ^2)\ \ ,\ \ \ \  \a  \ll 1\ ,
\label{masp}
\eeq
\beq
 \a =\tan\Theta =-{e q\over 4\pi }\ .
\label{dejo}
\eeq
In particular, the mass difference between a dyon and a monopole is given by
\beq
\Delta E\equiv M_{\rm dyon}- M_{\rm mon}\cong \ha |aq \a |\ .
\label{edio}
\eeq
The Coulomb potential has a constant piece at infinity, related to the energy associated
with the dyon electric field.
For the analysis in  Section 3,
it is convenient to write it as
\beq
A^0={\vec r.\vec\s\over 2r} V \hat q\ ,\ \ \ \ V=-{e a\over 4\pi } G(x)\cos\Theta \ .
\label{resz}
\eeq
where $\hat q$ represents the dyon charge operator.

\subsection{Charge conservation }
Let $\Phi =\Phi_b \s^b$ be the Higgs field which has a constant
expectation value at infinity,
and let  $\Psi_n$ stand for the other scalar fields.
We choose the gauge $A_0=0$.

The Lagrangian is invariant under the transformation
$$
\delta \Phi=0\ ,\ \  \ \delta \Psi_n =i [\epsilon  \Phi,\Psi_n]\ ,\ \ \
\delta A_l=\epsilon \p_l \Phi+i[\epsilon \Phi, A_l]\ .
$$
Consider now  $\epsilon =\epsilon (t)$ as an arbitrary function of
time. The Lagrangian is no longer invariant but the variation of
the Action with respect to $\epsilon$ must be zero because of the
equation of motions. We thus find a conserved quantity $Q$: \beq
Q=\int d^3 r\ {\rm Tr}\big( D_l \Phi \p_t A_l+ i \sum_n \Phi
[\Psi_n ,\p_t \Psi_n]\big) \ . \label{conser} \eeq One can indeed
independently check that $\p_t Q=0$ by using the field equations.
Note that $Q$ is given by an integral of the sum of two densities,
the ``charge density of the scalar fields'' $\Psi_n$, that is
$i{\rm Tr} \sum_n \Phi [\Psi_n ,\p_t \Psi_n]$ plus the ``charge
density of the dyon'', that is ${\rm Tr}( D_l \Phi \p_t A_l )$. A
similar proof for charge conservation is given in \cite{bala} for
the case of $SU(2)$ Yang-Mills theory with an isodoublet of Dirac
fermions.

Using the field equation
$$
D_l\p_t A_l=i \sum_n [\Psi_n,\p_t \Psi_n]+i[\Phi,\p_t \Phi]\ ,
$$
the charge can be written as
$$
Q=\int_{\Sigma } d\vec\Sigma \ {\rm Tr}[\Phi \p_t \vec A]\ ,
$$
where $\Sigma $ is a closed surface at infinity.

Thus $Q$ is interpreted
as the total charge (apart from a constant factor, equal to the expectation
value of the Higgs field at infinity) since it
is measured by the
total electric flux in the spontaneously broken theory.
Quantum mechanically, if the initial state is an eigenstate of $Q$, also the final
state must be an eigenstate of $Q$, since $Q$ commutes with the Hamiltonian.

\section{Scalar particles in the adjoint representation}

\subsection{Covariant Equations }

Let us consider  a scalar particle in the dyon background.
The scalar particle is a quantum of the matrix valued scalar field $\Psi$:
we take an $SU(2)$ triplet
$\Psi =\Psi_a\sigma^ \a $. The equation of motion  is given by
\beq
  D ^2_\mu \cdot \Psi
= {1\over 4} e^2 \big[ \Phi , \big[\Phi , \Psi \big] \big]\ ,\
\label{tre}
\eeq
where $  D ^2_\mu \cdot \Psi \equiv  D_\mu \circ( D_\mu\circ \Psi )$,
$ D_\mu \circ \Psi \equiv \partial_\mu \Psi -i e[ A_\mu,\Psi ]$.

We look for solutions of eq.~(\ref{tre}) of the form: $\Psi(\vec r,t)=e^{-iEt}\psi (\vec r) $,
and $\psi (\vec r) $ is
interpreted as the stationary wave function of the scalar particle.

It is convenient to choose the following  basis for $SU(2)$ Lie
algebra:
\beq
\hat \alpha= u_i \s_i\ ,\ \ \ \ \hat\b =v_i\s_i\ ,\ \ \ \hat\g =n_i \s_i\ ,
\eeq
\beq
 \ u_i=\delta_{3i}-n_3n_i\ ,\ \ \ \
v_i=\epsilon_{3ji} n_j \ ,\ \ \ \    n_i={r_i\over r}\ .\
\eeq
 Note that ${\vec u\over n_T}\ ,\  {\vec v\over n_T}, \ \vec n $, with $ n_T^2\equiv 1-n_3^2$
represent an orthonormal
frame for vectors in ${\bf R}^3$, $\vec u . \vec v=\vec u . \vec n=\vec n . \vec v=0$,
$\vec n^2=1$, $\vec u ^2=\vec v^2= n_T^2$.
It follows that ${\rm Tr}\ \hat\a\hat\b =  {\rm Tr}\ \hat\a\hat\g =  {\rm Tr}\ \hat\g\hat\b =0$,
${1\over 2} {\rm Tr}\ \hat\a ^2 = {1\over 2} {\rm Tr}\ \hat\b ^2  = n_T^2$,
${1\over 2} {\rm Tr}\ \hat\g ^2=1$.

The matrices $\hat\a ,\hat\b ,\hat\g $ obey the commutation relations:
$$
[\hat\a ,\hat\g]=2i\hat\b\ ,\ \ [\hat\b , \hat\g ]=-2i\hat\a , \ [\hat\a,\hat\b]=-2i n_T^2\hat\g\ ,
$$
and
$$
[\hat\a_\pm ,\hat\g ]=\pm 2\hat\a_\pm\ ,\ \ \ \ \hat\a_\pm =\hat\a\pm i\hat\b\ .
$$
The matrix valued wave function $\psi (\vec r)  $ can be decomposed as
\beq
\psi (\vec r) =F_+ (\vec r) \ \hat\a_ - +  F_- (\vec r)\ \hat\a_+ + F_0 (\vec r) \hat\g\ .
\label{onda}
\eeq
The component $F_0 (\vec r)$ multiplying $\hat\g $ represents a neutral component, whereas the
components $F_\pm $
of $\hat\a\mp i\hat\b $ represent charge $e$ and charge $-e$ components,
respectively.
This is clear by a  gauge transformation so that
\beq
U^{-1}\hat\g\ U=\sigma_3 \ ,\ \ \ \
U^{-1} (\hat\a \pm i\hat\b ) U= -2\sin\theta \ e^{\pm i\varphi } \sigma_\mp\ ,
\label{zzz}
\eeq
where
\beq
U(\vec n)=\cos{\theta\over 2} +i\sin
{\theta\over 2}\ {\epsilon_{3ij} \s_i n_j\over \sin\theta }\ .
\label{zuum}
\eeq
In this gauge, the charge operator is $\hat Q={e\over 2}\ \s_3$, acting on a state $\Psi $
by $\hat Q \cdot \Psi \equiv [\hat Q, \Psi ]$.

Now we evaluate
$\vec D^2\cdot \psi  $. With the definitions (1.4), and further defining for short
$\Omega \equiv
\vec\p ^2-{1\over
r^2}(1+K^2)-{2\over r}{n_3\over n_T^2}u_j\p_j$, we find:
\beq
\vec   D^2 \cdot \big( \hat\g \ F_0 (\vec r) \ \big)=  \hat \g  \ \bigg( \vec\p
^2-{2K^2\over r^2}\bigg) F_0
 +\hat\a \ \bigg( {2K \over r n_T^2}u_j\p_j  \bigg) F_0 +
  \hat\b \ \bigg(  {2K \over r n_T^2}v_j\p_j  \bigg) F_0 \ .
\label{udieci}
\eeq
 \beqa
\vec   D^2 \cdot \big[ (\hat\a \pm i\hat\b )F_\mp \big] &=&
(\hat\a \pm i\hat\b )\bigg( \Omega
\pm {2\over r}  {n_3\over n_T^2}iv_j\p_j \bigg ) F_\mp
\non\\
&+& \hat\g\ {2K\over r}
\bigg( \mp iv_j\p_j-u_j\p_j+
{2n_3\over r}\bigg)F_\mp \ .
\label{qqf}
\eeqa

\subsection{Harmonic analysis }

To solve for the angular dependence,
it is convenient to use spherical coordinates $r, \theta ,\vp $.
We have:
\beq
u_j \p _j = - {1\over r} \sin\theta\ \p_\theta\ ,\ \ \ \ \ i v_j \p _j  = i{1\over r}\p_ \vp\ .
\label{esfe}
\eeq
Asymptotically at $r\to \infty $,
we have $K\to 0$ and in this case  there is decoupling of charges $(+,-,0)$  in (\ref{udieci}) and (\ref{qqf}), as expected.
Consider the monopole case $\Theta=0$.
Charged particles have mass $m$, with $m=ea$.
Define
\beq
F_\pm (\vec r)={e^{\pm i\vp }\over \sin\theta }h_\pm (\vec r)\ .
\label{ffpm}
\eeq
We find  that for $r\gg m^{-1}$ eqs.~(\ref{tre}), (\ref{qqf}) give
$$
\bigg( {1\over r^2} \p_r r^2  \p_r +{1\over r^2\sin\theta } \p_\theta (\sin\theta \p_\theta )
+ {1\over r^2\sin^2\theta }
\big( \p_\varphi \mp i (1-\cos\theta )\big) ^2
$$
$$
+
E^2 - (m-{1\over r})^2 \bigg) h_\mp (\vec r)=0\ .
$$
This is the covariant equation for a scalar particle of charge $\mp e$
moving in a $U(1)$
point-like Dirac monopole background. In addition,
there is a Coulomb-like potential due to the long range
tail of the Higgs field.

Let us now consider the complete equations in all the space.
In the Appendix we report
the harmonic analysis in full detail.
The angular dependence is solved by setting
$$
F_0(\vec r)= \phi_0(r)\ Y_{lm}(\theta,\vp )\ ,\ \
$$
$$
h_+ (\vec r)= {1\over l_0}\ \phi_+ (r)\ Z_{lm}^+(\theta,\vp )\ ,\ \ \ \
h_- (\vec r)={1\over l_0 }\ \phi_- (r)\ Z_{lm}^-(\theta,\vp )\ ,\ \ \
$$
$$
 l_0\equiv\sqrt{l(l+1)}\ ,
$$
where $ Y_{lm}(\theta,\vp )$ are the standard spherical harmonics,
and $Z_{lm}^\pm $ can be expressed in terms of them:
\beq
Z_{lm}^{\pm}= {e^{\mp i\vp }\over \sin\theta } (\sin\theta \p_\theta\mp i \p_\vp )Y_{lm}\ .
\eeq
Note that the angular expansion of the charged components $h_{\pm}$ begins with $l=1$.

Using the results of the Appendix,
we get the following system of coupled differential equations for $\phi_0(r), \ \phi_\pm (r)$ :
  \beq
 \bigg( -{1\over r^2}\p_r r^2\p_r
+{2K^2\over r^2}+{1\over r^2} l(l+1)\bigg) \phi_0
+ {2K l_0\over r^2   }\big(  \phi_+  +  \phi_- \big)
 =E^2 \phi_0\ ,
\label{aazi}
\eeq
\beqa
  & \big[ -{1\over r^2}\p_r r^2\p_r
 +{1\over r^2}(-1+K^2)+{1\over r^2}l(l+1) \big]\phi_+
 + {K l_0\over r^2}   \phi_0
\non\\
& =\big[\big( E  -e V \hat q )^2  - e^2 a^2  G^2 \big] \phi_+\ ,
\label{aapi}
\eeqa
 \beqa
  & \big[ -{1\over r^2}\p_r r^2\p_r
 +{1\over r^2}(-1+K^2)+{1\over r^2} l(l+1) \big]\phi_-
 + {K l_0\over r^2 } \phi_0
\non \\
& =  \big[\big( E  + e V \hat q )^2  - e^2 a^2  G^2 \big] \phi_-\ ,
\label{aapp}
\eeqa
where $V$ was defined in eq.~(\ref{resz}).

\subsection{ Collective coordinates and equations of motion}

When a process of charge transfer
occurs, there is an important back reaction in the dyon background.
This effect can be taken into account by incorporating into the dynamics
the  collective coordinate $\chi $ associated to global $U(1)$ gauge transformations.
This ``quantum rotator" degree of freedom ensures charge conservation.
The  dyon charge operator is written as $\hat q=g \tan\Theta = -i e \p_\chi $.

Consider a system consisting of a scalar particle and a dyon with
total charge equal to zero. The wave function can be written in
the form \beq \psi = \left( \matrix{  \phi_+\ e^{-i\chi } \cr
\phi_0 \cr \phi_- \ e^{i\chi } }\right)\ . \eeq In general, this
represents a mixture of  monopole and neutral scalar, with dyons
and scalars of charges $(+-)$ and $(-+)$. When the system makes a
transition from  a monopole to a dyon, the field component $A_0$
is turned on. In addition, the energy of the scalar is reduced
into an amount equivalent to the mass difference between  the
dyon and the monopole. In what follows we will assume that $\a
={e^2\over 4\pi }$ is small. If $\a $ is not small, other back
reaction effects become important and this semiclassical analysis
is not applicable (for $\a \ll 1$, one may approximate
$\sin\Theta \cong \tan\Theta $, $\cos\Theta=1+O(e^4)\cong 1$).

The equations of motion (\ref{aazi})-(\ref{aapp}) can be written as
\beq
H \psi =0\ ,
\eeq
where
$$
H=\bigg(\hat P  + \big( E +
 ms  G
i\p_\chi \big) ^2 -m^2 G^2   \bigg)
{\bf 1}_{3\times 3}
+{l_0\over r^2} \hat M\ ,
$$
$$
\hat M=   \left( \matrix{ 0 &Ke^{-i\chi } &0 \cr
                      Ke^{i\chi } &0  &K e^{-i\chi }\cr
                        0 &K e^{i\chi } &0 }
\right)\ ,\ \ \ s\equiv\sin\Theta\ ,
$$
$$
\hat P= -{1\over r^2}\p_r r^2\p_r + {1\over r^2}l(l+1)
 +{2K^2\over r^2}+{1\over r^2}(1+K^2)\p_\chi^2\ .
$$
Then the equations of motion for the neutral system take the form
 \beq
 \bigg( -{1\over r^2}\p_r r^2\p_r
+{2K^2\over r^2}+{1\over r^2} l(l+1)\bigg) \phi_0
+ {2K l_0\over r^2   }\big(  \phi_+  +  \phi_- \big)
 =E^2 \phi_0\ ,
\label{bbzi}
\eeq
\beqa
  & \big[ -{1\over r^2}\p_r r^2\p_r
 +{1\over r^2}(-1+K^2)+{1\over r^2}l(l+1) \big]\phi_+
 + {K l_0\over r^2}   \phi_0
\non\\
& =\big[\big(E - ms   G  \big)^2 - m^2  G^2\big] \phi_+\ ,
\label{bbpi}
\eeqa
 \beqa
  & \big[ -{1\over r^2}\p_r r^2\p_r
 +{1\over r^2}(-1+K^2)+{1\over r^2} l(l+1) \big]\phi_-
 + {K l_0\over r^2 } \phi_0
\non \\
& =\big[\big(E  - ms  G  \big)^2 - m^2  G^2\big] \phi_- \ .
\label{bbpp}
\eeqa
 The incorporation of the collective coordinate $\chi $ produces
the flip of sign of the dyon charge in the equations for $\phi_+ $ and $\phi _-$, required by charge conservation.
As a result, the Coulomb potential of the dyon-scalar is always attractive for both $\phi_+ $ and $\phi _-$.

In terms of the variables $\phi_{\rm a}= \phi_+  - \phi_- $
and $\phi_{\rm s} =\phi_+ + \phi_- $, we have a decoupled equation
\beqa
  & \big[ -{1\over r^2}\p_r r^2\p_r
 +{1\over r^2}(-1+K^2)+{1\over r^2} l(l+1) \big]\phi_{\rm a}
\non \\
& =\big[ (E -  m s  G)^2   - m^2  G^2 \big] \phi_{\rm a}\ ,
\label{aaee}
\eeqa
and a coupled system of ordinary differential equations for $\phi_{\rm s}$ and $\phi_0 $,
     \beqa
  & \big[ -{1\over r^2}\p_r r^2\p_r
 +{1\over r^2}(-1+K^2)+{1\over r^2} l(l+1) \big]\phi_{\rm s}
 + {2K l_0\over r^2 } \phi_0
\non \\
& =\big[\big(E  - m s  G \big)^2 - m^2 G^2\big]
 \phi_{\rm s}\ ,
\label{aacc}
\eeqa
   \beq
 \bigg( -{1\over r^2}\p_r r^2\p_r
+{2K^2\over r^2}+{1\over r^2} l(l+1)\bigg) \phi_0
+ {2K l_0\over r^2   }  \phi_{\rm s}
 =E^2  \phi_0\ .
\label{ssgg}
\eeq

In the Section 3.4, we will see that eq. (\ref{aaee}) describes  bound states.
The wave function
for such bound state is of the form
$|\psi \rangle =|{\rm dyon}_+\rangle |{\rm scalar}_-\rangle
- |{\rm dyon}_-\rangle |{\rm scalar}_+\rangle $,
i.e. there is quantum entanglement
of charges with a dyon state.
Eqs. (\ref{aacc}), (\ref{ssgg}) can be used to describe the  dynamics of  a scattering process in which
a neutral scalar particle scatters off a monopole and gives rise to an outgoing state
which is a mixture of a neutral scalar and a $+/-$ charged scalar entangled with a $-/+$ charged dyon.

\subsection{Bound states}

Here we study the solutions of the equation (\ref{aaee}) for $\phi_{\rm a} $. In terms of the radial coordinate
$x=(m\cos\Theta )\ r$, it takes the form
\beqa
  & \big[ -{1\over x^2}\p_x x^2\p_x
 +{1\over x^2}(-1+K^2(x))+{1\over x^2} l(l+1) \big]\phi_{\rm a}
\non \\
& =\big[ \E^2 - 2\E \a  G(x)   -   G^2(x) \big] \phi_{\rm a}\ ,
\label{aaff}
\eeqa
$$
\E\equiv {E\over m\cos\theta }   \ ,\ \ \ \ \a=\tan\Theta={e^2\over 4\pi }\ .
$$
This is a Schr\" odinger-type equation with a potential which is an attractive Coulomb potential
at large distances, where $G\cong 1-{1\over x}$
(see eq.~(\ref{cou})~),
and therefore it may admit bound state solutions.

A closed analytic solution of this differential equation is not known.
We will solve it by using two independent methods:
a) Analytic method;
b) Numerical method.

Let us begin with the analytic method. We approximate
the potential by a simpler function,
replacing $K^2$ and $G$ by $K_0^2$, $G_0$ defined as follows (cf. eqs.~(\ref{cou}),(\ref{reg})~):
$$
K_0^2(x)=0\ ,\ \ \ \ \ G_0(x)=1-{1\over x}\ \ \ ,\ \ x> 1  \ ,
$$
$$
K_0^2(x)=1-{x^2\over 3}\ ,\ \ \ \ \ G_0(x)=0\ \ \ ,\ \ x\leq 1\ .
$$

\medskip

\noindent {\it  Exterior solutions:} For $x>1$ the equation is
 \beq
\big[   -{1\over x^2}\p_x x^2\p_x +{l(l+1)\over x^2} -{2\over x}(\a  \E+1)
+k^2\big]\phi_{\rm a}  = 0
\label{zpa}
\eeq
with
\beq
 k=\sqrt{1+2\a\E- \E^2}\ .\ \
\label{kaka}
\eeq
Note that the Coulomb potential in eq. (\ref{zpa}) has two contributions,
one coming from the interaction with
the Higgs field,
the other from  $A_0$. For $\a \ll 1$, the Coulomb potential due to the
Higgs field is dominant.

Solutions with real $k$ represent bound states.
The solution which vanishes at infinity is the confluent hypergeometric function:
\beq
 \phi_{\rm a}^{ex}=  x^l e^{-k x} \Psi (a,2+2l,-2 k x)\ ,\ \ \ \  a=1+l-{\a  \E+1\over k}\ .
\label{afu}
\eeq
Equation (\ref{zpa}) is formally the same as the Schr\"odinger equation for the Hydrogen atom.
Thus, for bound states with wavefunctions having support at $ x \gg 1$,
the energy eigenvalues are approximately determined by the hydrogen atom formula.
This formula follows by demanding that the wave function be regular at the origin, which
amounts to say that the parameter $a$ is an integer $\leq 0$.
Thus in this case the eigenvalues are
given by
\beq
  n= n_0+l+1= {\a  \E+1\over k}\ ,\ \ \ n_0=0,1,2,...
\eeq
or
\beqa
\label{awer}
\E &=& {\a (n^2-1)\pm n \sqrt{n^2-1} \sqrt{\a^2+1} \over n^2+\a^2}\ ,\ \
\\
&\cong &  \pm \sqrt{1-{1\over n^2} }+\a (1-{1\over n^2})\ ,\ \ \ \a \ll 1\ .
\non
\eeqa
We shall see   that (\ref{awer}) is a very good approximation for all eigenstates with $l>1 $.

Note that there are also eigenvalues of negative energies. The
negative energy eigenstates should be interpreted in terms of the
antiparticle of opposite charge (thus $\alpha \to -\alpha $, see
eq.~(\ref{dejo})~) and positive energy. Reversing the sign of the
energy  has the same effect in the equation as reversing the sign
of $\alpha $. These states no longer correspond to the neutral
system under discussion, but to a system of total electric charge
equal to $2e$ (or $-2e$). They are bound to the dyon because the
Coulomb potential is still attractive thanks to the Higgs
contribution (i.e. $\a \E<0 $ but $1+\a \E >0$).
  Here we will discuss only eigenvalues of positive energies.

Note that there are bound states
 even in the limit  $\a \to 0$. The origin of the binding force in this case is the Higgs field.

The binding energy is represented by
$$
k^2= 1+2\a\E-\E^2= ( \E_\infty^+ -\E ) (\E - \E_\infty ^-)\ ,
$$
with
$$
\E_\infty^\pm =\a \pm \sqrt{1+\a^2}\ .
$$
Here $\E_\infty ^+$ represents the asymptotic mass.


\medskip

\noindent {\it  Interior solutions:} For $x<1$ the equations are
\beq
\big[   -{1\over x^2}\p_x x^2\p_x
  +{l(l+1)\over x^2}-k_{\rm in}^2\big]\phi_{\rm a}  = 0\ ,
\label{zpi}
\eeq
with
$$
 \ k_{\rm in}^2=
\E ^2 +{1\over 3}\ .
$$
The solution that is regular at $r=0$ is given by
  \beq
\phi_{\rm a}^{in}=c_0 J_{l+{1\over 2}}\big( k_{\rm in} x \big)\ .
\eeq
The energy eigenvalues follow from imposing continuity of the wavefunction and its first derivative
at $x=1$.
This gives the condition
\beq
\big( \p_x \phi_{\rm a}^{in}\  \phi_{\rm a}^{ex}- \phi_{\rm a}^{in}\ \p_x  \phi_{\rm a}^{ex}\big)\bigg|_{x=1}=0\ .
\eeq
The resulting eigenvalues (in terms of $k^2=1+2\a\E-\E^2$)
are given in Table 1.

\begin{table}[htbp]
\centering
\begin{tabular}{l|l|ccc}
$l$ & $n_0$ & $k_{\rm num}$ & $k_{\rm an}$ & $k_{\rm point}$  \\
 \hline
1 & 0 & 0.5724 & 0.5953 & 0.6023 \\
1 & 1 & 0.3937   &  0.4048 & 0.4089    \\
2 & 0 & 0.4076   & 0.4094  & 0.4089   \\
2 & 1 & 0.3076 &  0.3086    & 0.3086  \\
3 & 0 & 0.3086 & 0.3086 & 0.3086  \\
3 & 1 &  0.2476 & 0.2476  & 0.2476 \\
\end{tabular}
\parbox{5in}{\caption{Eigenvalues  for binding energy
for $\a =0.2$.   $k_{\rm num}$ are  the eigenvalues obtained by
numerical integration while  $k_{\rm an}$ are obtained by the
analytic method. The last column $k_{\rm point}$ are the
eigenvalues in the point-like limit, where the differential
equation is formally the same as the hydrogen atom equation
(using (\ref{awer})~). \label{tab:Eadjoint}}}
\end{table}

\begin{figure}[hbt]
\label{fig1} \vskip -0.5cm \hskip -1cm
\centerline{\epsfig{figure=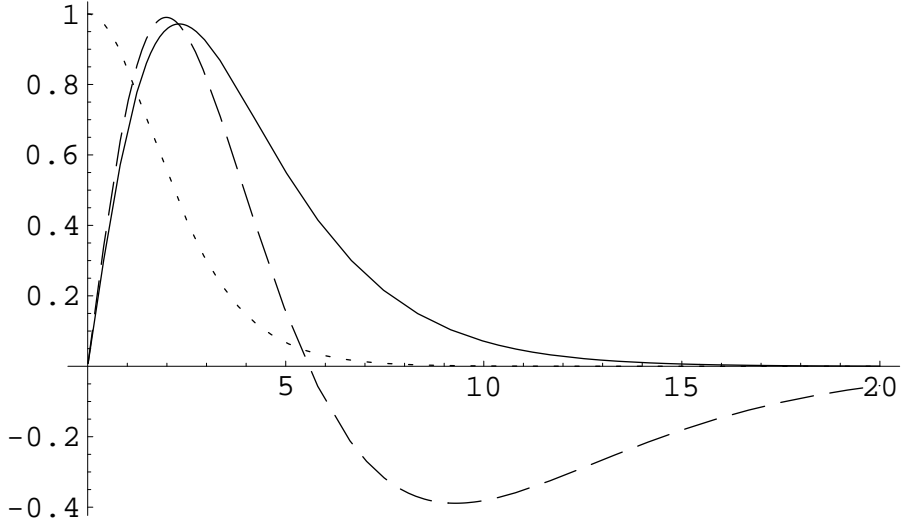,height=7truecm}}
\caption{\footnotesize Wave functions for the ground state
(solid line) and first excited state (dashed line) with $l=1$,
$n_0=0,1$, respectively, for $\a =0.2 $. The dotted line is the
plot of $K^2(x)$, which indicates the region of the dyon core.}
\end{figure}

\medskip

The numerical determination of the eigenvalues is performed by
integrating the differential equation from $x=0$ to some $x=x_{\rm max}\gg 1 $,
with regular boundary conditions at zero. The  energy eigenvalues are then
determined by the condition that the wavefunction approaches zero at $x=x_{\rm max}$
(equivalently, that the wave function approaches the asymptotic form (\ref{afu})~).
 The results are given in Table 1. Fig.~1 is a plot of the wave functions for the
ground state and the first excited level.

A final remark about the stability of the bound states. The theory contains also photons,
i.e. the quanta of the (far from the core) unbroken $U(1)$ gauge field, which are coupled
to our charged scalar particles. The coupling is proportional to the charge operator,
which has nonvanishing matrix elements between the states $\phi_{\rm a} $ and
the states $\phi_{\rm s} $.
Because $\phi_{\rm s} $ is coupled to the massless scalar $\phi_0 $,
this implies an instability of the bound state.
 The dominant decay channel of the fundamental state $\phi_{\rm a} $ is  by single photon emission, and subsequently conversion of $\phi_{\rm s} $ into a $\phi_0$ particle that can get to infinity.
A further discussion of this is given in Section 5.

\section{Scalar particles in the fundamental representation}

\subsection{Covariant Equations}
We will work
in the gauge where the charge operator is $\hat Q={e\over 2}\
\s_3$. Now the wave function $\psi (\vec r)$ is an isospinor, that
is an isospin doublet, and the charge operator acts on it in the
standard way as a $2\times 2$ matrix on an isospinor. In this
gauge, the space components of the covariant derivatives are \beqa
\hat D_l &\equiv & U^{-1}(\p_l -ieA_l)U \non\\
         &=& \p_l +{i\over 2r(1-n_3^2)}\lbrack K\ (u_l\vec v\cdot\vec\sigma +v_l\vec n_T\cdot\sigma )
             +v_l(1-n_3)\sigma_3 )\rbrack\ .
\eeqa
Here we use the notation of Section 3.1, with $U$ defined in eq.~(\ref{zuum}).

Notice that for $K\cong 0$ (i.e., far away from the core) one has
\beqa
\hat D_l^2 &=& (\p_l+{iv_l\over 2r(1+n_3)}\sigma_3)^2 \non\\
           &=& {1\over r^2}\p_r r^2\p_r+{1\over r^2\sin^2\theta }
 \bigg(\sin\theta\p_\theta (\sin\theta\p_\theta )+\big( \p_\vp+{i\over 2}(1-\cos\theta )\sigma_3\big)^2\bigg).
\non
\eeqa
This is the covariant Laplace operator for a scalar particle of charge $\pm \ha e$  moving in  a $U(1)$ Dirac monopole background.

The full equations can be written as: \beq -\hat D_l^2 \cdot\psi
=\bigg(-{1\over r^2}\p_r r^2\p_r +{K^2\over 2 r^2}-{1\over
4r^2}+{1\over r^2} \hat J^2_{1/2}\bigg)\psi +{K\over r^2 }M\cdot
\psi\ , \eeq
 where
\beq \hat J^2_{1/2}=\left(\matrix{\hat J^2_{+1/2}&0\cr 0&\hat
J^2_{-1/2}}\right) \ , \ \ \ M={1\over \sin\theta}\left( \matrix{
0 &e^{-i\vp }\nabla_{-1/2 } \cr -e^{i\vp }\bar \nabla_{1/2} &0}
\right)\ ,\ \ \eeq
\beq
  \psi= \left(\matrix{\psi_+ \cr \psi_-}
\right)\ . \eeq
Now \beq \hat J^2_{\pm 1/2}=-{1\over \sin^2\theta
}\nabla_{\pm 1/2}\bar\nabla _{\pm 1/2} \eeq with \beq \nabla_{\pm
1/2} =\sin\theta \p_\theta \mp i\p_\vp + \ha (1-\cos\theta )\ ,
\eeq \beq \bar \nabla_{\pm 1/2} =\sin\theta \p_\theta \pm i\p_\vp
-\ha (1-\cos\theta )\ . \eeq

\subsection{Harmonic analysis}

 Let
$Z_{lm}^{(\pm\ha )}$ be an eigenfunction of the operator
\beq
\tilde L^2_{\pm} =\hat J^2_{\pm 1/2}-1/2+1/4\ ,
\eeq
In particular consider
$$
\tilde L^2_{+}     Z_{lm}^{(\ha )}=l(l+1)  Z_{lm}^{(\ha )}\ .
$$
Explicit expressions for these eigenfunctions are given in ref.~\cite{yang,kazam}.
The eigenvalues are now $l=\ha ,\ha+1,\ha+2, \dots $\
Note that $\hat J^2_{\pm 1/2} =(l+1/2)^2$.
It is convent to fix a phase convention by defining
\beq
   Z_{lm}^{(-\ha )}\equiv -{1\over\mu }\ {e^{i\vp} \over \sin\theta }\bar \nabla_{1/2}   Z_{lm}^{(\ha )} \ ,
\ \ \ \ \mu\equiv  l+\ha\ .  \label{mu}
\eeq
It is indeed easy to verify that
\beq
   Z_{lm}^{(\ha )} = {1\over\mu } \ {e^{-i\vp} \over \sin\theta } \nabla_{-1/2 }   Z_{lm}^{(-\ha )}\ ,\ \ \ \
\tilde L^2_{-} Z_{lm}^{(-\ha )}=l(l+1)  Z_{lm}^{(-\ha )}\ .
\eeq
The angular dependence is then solved by writing
$$
\psi_+=\eta_+ (r) Z_{lm}^{(\ha )}(\theta,\vp )\ ,\ \ \ \ \psi_- = \eta_- (r) Z_{lm}^{(-\ha )}(\theta,\vp )\ .\
$$

Next, we compute $\hat D_0^2 \cdot\Psi $, $\hat D_0=U^{-1} (\p_0-\ha i e A_0) U$.
Using $U^{-1}\g U=\s_3 $, we find
\beq
\hat D_0^2\cdot \Psi =-(E- \ha e\hat q V \ \s_3)^2\Psi \ .
\label{coulo}
\eeq
Now we consider a system of vanishing total charge.
To incorporate charge conservation, as in Section 3.2 we write the dyon charge operator
as
 $\hat q=-i e\p_\chi $, and the wave function as
\beq
\psi = \left( \matrix{\eta_+\ e^{-\ha i\chi } \cr \eta_- \ e^{\ha i\chi } }\right)\ .
\eeq
The net effect is that the cross term in the square appearing on the right hand side  of (\ref{coulo})
(which contains the Coulomb interaction)
has the same sign for the upper and lower components, i.e. it is an attractive potential for both components.

\subsection{Case of the $N=2$ Hypermultiplet }

{}From the point of view of $N=1$ supersymmetry, the $N=2$ vector multiplet
contains a $N=1$ vector multiplet and a chiral multiplet $\Phi $.
In this Section we add  to the $N=2$ pure Yang-Mills theory  a hypermultiplet in the fundamental representation of $SU(2)$.
The  hypermultiplet contains two chiral superfields $Q$ and $\tilde Q$, which couple
to $\Phi $ by a term $W=\tilde Q \Phi Q$.
  The fundamental scalars  will get  a mass due to the coupling to the Higgs field.
To simplify the discussion, here we will not add
an independent mass term $M \tilde Q Q$.
 In the next subsection, we will consider scalar particles
with an arbitrary mass parameter.

Let us consider a neutral system of dyon and scalars with charges $(\ha ,-\ha )$ and $(-\ha, \ha )$.
Taking into account the coupling to the Higgs field, we  get the
 following equations for $\eta_+ (r)$ and $\eta_- (r)$ ($\mu$ being defined in (\ref{mu})~):
\beq
\big[   -{1\over r^2}\p_r r^2\p_r +{K^2\over 2 r^2}+{1\over r^2} (\mu^2-\ha )\big]\eta_+ + {\mu K\over r^2} \eta_- =
\big[(E-\ha ms  G)^2-\four m^2 G^2\big]\eta_+\ ,
\label{zzu}
\eeq
  \beq
\big[   -{1\over r^2}\p_r r^2\p_r +{K^2\over 2 r^2}+{1\over r^2} (\mu^2-\ha )\big]\eta_- + {\mu K\over r^2} \eta_+ =
\big[(E-\ha ms  G)^2-\four m^2 G^2\big]\eta_-\ ,
\label{yyu}
\eeq
 where $s=\sin\Theta $.
 Let us now define $\eta_{\rm s} =\eta_++\eta_- $\ ,\   $\eta_{\rm a} =
\eta_+ - \eta_- $.  They satisfy the decoupled equations
\beq
\big[   -{1\over r^2}\p_r r^2\p_r +{K^2\over 2 r^2}+{1\over r^2} (\mu^2-\ha )\big]\eta_{\rm s} +
{\mu K\over r^2} \eta_{\rm s} =
\big[(E-\ha ms G)^2-\four m^2G^2\big]\eta_{\rm s}\ ,
\label{zzuu}
\eeq
  \beq
\big[   -{1\over r^2}\p_r r^2\p_r +{K^2\over 2 r^2}+{1\over r^2} (\mu^2-\ha )\big]\eta_{\rm a} -
{\mu K\over r^2} \eta_{\rm a} =
\big[(E-\ha ms G)^2-\four m^2 G^2\big]\eta_{\rm a}\ .
\label{yyuu}
\eeq
Note that the equations for $\eta _{\rm s}$ and $\eta _{\rm a}$
are formally the same under the exchange $\mu \to -\mu $, so they can be
investigated on the same footing. In what follows we consider the $\eta_{\rm s}$ equation
(\ref{zzuu}), wherefrom we obtain the solutions for    $\eta_{\rm a}$ by flipping the sign
of $\mu $.

In terms of $x=(m \cos\Theta )\ r$, the equation  (\ref{zzuu}) for $\eta_{\rm s}$ reads
$$
\big[   -{1\over x^2}\p_x x^2\p_x +{K^2(x)\over 2 x^2}+{1\over x^2} (\mu^2-\ha )\big]\eta_{\rm s} +
{\mu K(x)\over x^2} \eta_{\rm s}
$$
\beq = \big[\E ^2 -  \a \E G(x)- \four G^2(x) \big]\eta_{\rm s}\ ,
\label{hyyuu} \eeq
$$
\E={E\over m \cos\Theta }\ ,\ \ \ \a =\tan\Theta\ .
$$

At $x\gg 1$, the differential equation (\ref{hyyuu}) takes the
form \beq
   -{1\over x^2}\p_x x^2\p_x \eta_{\rm s}+\bigg({l(l+1)\over x^2}  -{2\a \E +1\over 2x}
+k^2\bigg)\eta_{\rm s}  = 0\ ,
\label{zzv}
\eeq
with
$$
k=\sqrt{ \four + \a \E  -\E^2  } \ .\
$$
The solution which vanishes at infinity is the confluent hypergeometric function:
\beq
 \eta_{\rm s}\cong  x^l\ e^{-k x} \Psi ( a,2+2l,-2 k x)\ ,\ \ \ \
\label{afue}
\eeq
$$
 \  a=1+l - {2\a  \E +1\over 4k  } \ .
$$
The approximate energy eigenvalues are  then determined by the condition
\beq
   n_0+l+1= {2\a  \E+1\over 4\sqrt{ \four + \a \E  -\E^2  }}\ ,\ \ \ n_0=0,1,2,...
\label{atomm}
\eeq
Some values are given in Table 2.
Note that energy eigenstates obtained in this approximation are the same for $\eta_{\rm s}$ and $\eta_{\rm a}$, since at $x\gg 1$ $\eta_{\rm a}$ satisfies the same eq.~(\ref{zzv}).
This point-like approximation is already good for states with $l=3/2$ and becomes better for states with high angular momentum, which are farther from the core.

 \begin{table}[htbp]
\centering
\begin{tabular}{l|l|ccc}
l & $n_0$ &
$k_{\rm num}$(s) & $k_{\rm num}$(a) & $k_{\rm point}$  \\
 \hline
1/2 & 0 & 0.1921 & 0.2245 & 0.2045  \\
1/2 & 1 &  0.1188 & 0.1310 &  0.1238    \\
3/2 & 0 &  0.1236 & 0.1241 & 0.1238  \\
3/2 & 1 &  0.08848 & 0.08881 & 0.08863    \\
5/2 & 0 & 0.08863 & 0.08863 & 0.08863  \\
5/2 & 1 &  0.069005 & 0.069005 & 0.069005 \\
\end{tabular}
\parbox{5in}{\caption{Eigenvalues for binding energies   $k_{\rm num}$(s,a)
for $\eta_{\rm s}$ and $\eta_{\rm a}$ using $\a =0.2$. $k_{\rm
point}$ are the approximate eigenvalues given by the analytic
expression
 (\ref{atomm}).
\label{tab:Efunda}}}
\end{table}

\begin{figure}[hbt]
\label{fig2} \vskip -0.5cm \hskip -1cm
\centerline{\epsfig{figure=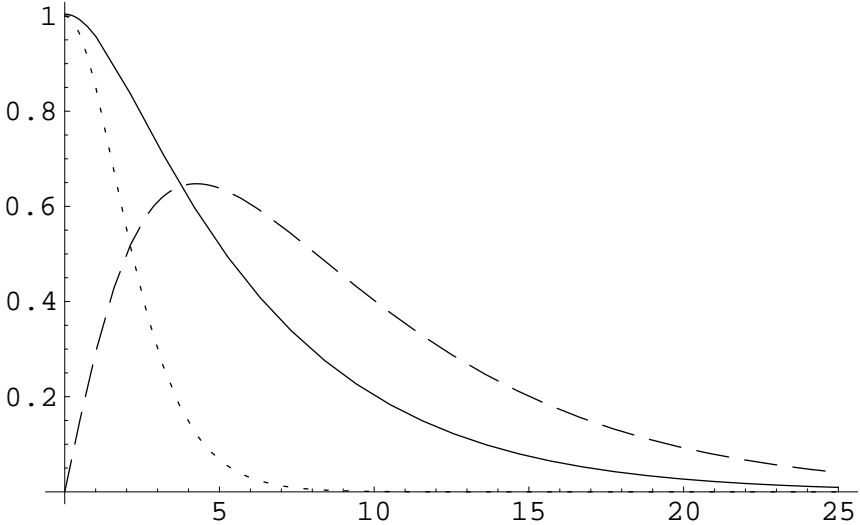,height=7truecm}}
\caption{\footnotesize Wave function for the ground state
$\eta_{\rm a}$ (solid line) and for $\eta_{\rm s}$ (dashed line)
with  $l=1/2$, $n_0=0$ (we use $\a =0.2 $). The dotted line is a
plot of $K^2$, indicating the region of the dyon core.}
\end{figure}


At $x\ll 1$, the equation is
\beq
   -{1\over x^2}\p_x x^2\p_x
  \eta_{\rm s}+\bigg({\mu (\mu+1)\over x^2}-k_{\rm in}^2\bigg)\eta_{\rm s}  = 0\ ,
\label{nzzv} \eeq with
$$
\ k_{\rm in}^2=  \E ^2 +{ (1+\mu ) \over  6}\ ,\ \ \ \ \ \mu=l+\ha \ \ .
$$
The solutions are Bessel functions.

In the case of  $\eta_{\rm a}$, the ``centrifugal barrier''
for $x\ll 1$ is   $ {\mu (\mu-1)\over x^2} $.
As a result, there is no barrier for the ground state with $\mu= 1$, and
the corresponding wave function does not vanish at the origin (see fig.~2).

\medskip

 The numerical determination of the eigenvalues is performed as in Section  3 by
integrating the differential equation from $x=0$ to some $x=x_{\rm max}\gg 1 $,
with regular boundary conditions at zero. The  energy eigenvalues are then
determined by the normalizability requirement that the wavefunction approaches zero at $x=x_{\rm max}$.
 The results are given in Table 2.

\medskip

\subsection{Massive scalars in $N\leq 1$ without coupling to Higgs}

In the systems of Section  3.4 and 4.3, for small $\a $ the Coulomb attraction is dominated by
the Higgs field. This interaction is dictated by $N=2$ supersymmetry.
As a result, the bound states are not very sensitive to the value of an $\a <1 $.
It is of interest to investigate systems with less supersymmetry,
to see the effect of changing the values of the coupling.
Here  we will consider a scalar field in the fundamental representation
with an arbitrary mass term and no coupling to the Higgs field.

Let us consider a neutral system of dyon and these scalars with charges $(\ha ,-\ha )$ and $(-\ha, \ha )$.
In a similar way as in Section  4.3,
 we find the following equations for $\eta_+ (r)$ and $\eta_- (r)$~:
\beq
\big[   -{1\over r^2}\p_r r^2\p_r +{K^2\over 2 r^2}+{1\over r^2} (\mu^2-\ha )\big]\eta_+ + {\mu K\over r^2} \eta_- =
\big[(E-\ha ms  G)^2-M^2\big]\eta_+\ ,
\label{qzzu}
\eeq
  \beq
\big[   -{1\over r^2}\p_r r^2\p_r +{K^2\over 2 r^2}+{1\over r^2} (\mu^2-\ha )\big]\eta_- + {\mu K\over r^2} \eta_+ =
\big[(E-\ha ms  G)^2-M^2 \big]\eta_-\ ,
\label{qyyu}
\eeq
where  we have  introduced a
mass $M$.
Thus, again the presence of $\eta_+$ turns on the other component
$\eta_- $ (and viceversa), so the bound state is an entangled quantum state. The equations are decoupled in terms of
$\eta_{\rm s} =\eta_++\eta_- $\ ,\   $\eta_{\rm a} =
\eta_+ - \eta_- $, which satisfy
\beq
\big[   -{1\over r^2}\p_r r^2\p_r +{K^2\over 2 r^2}+{1\over r^2} (\mu^2-\ha )\big]\eta_{\rm s} +
{\mu K\over r^2} \eta_{\rm s} =
\big[(E-\ha ms G)^2-M^2\big]\eta_{\rm s}\ ,
\label{qzzuu}
\eeq
  \beq
\big[   -{1\over r^2}\p_r r^2\p_r +{K^2\over 2 r^2}+{1\over r^2} (\mu^2-\ha )\big]\eta_{\rm a} -
{\mu K\over r^2} \eta_{\rm a} =
\big[(E-\ha ms G)^2-M^2 \big]\eta_{\rm a}\ .
\label{qyyuu}
\eeq
Introducing the rescaled radial coordinate $x=(m \cos\Theta )\ r$, the equation for $\eta _{\rm s}$ becomes
$$
\big[   -{1\over x^2}\p_x x^2\p_x +{K^2(x)\over 2 x^2}+{1\over x^2} (\mu^2-\ha )\big]\eta_{\rm s} +
{\mu K(x)\over x^2} \eta_{\rm s}
$$
\beq
=
\big[(\E -\ha  \a G(x))^2-{\cal M}^2 \big]\eta_{\rm s}\ ,
\label{qqyyuu}
\eeq
$$
\E={E\over m \cos\Theta }\ ,\ \ \ {\cal M}={M\over    m \cos\Theta }\ .
$$
In the region $x\gg 1$, this differential equation reduces to
\beq
   -{1\over x^2}\p_x x^2\p_x \eta_{\rm s}  +\bigg({1\over x^2} (\mu^2-\ha-\four \a^2) -{\a \E\over x}
+k^2\bigg)\eta_{\rm s}  = 0\ ,
\label{qzzv}
\eeq
with
$$
k=\sqrt{{\cal M}^2- (\E -\ha \a ) ^2}\ .\ \
$$
The solution which is normalizable at infinity is  the following confluent hypergeometric function:
\beq
 \eta_{\rm s}\cong  x^{b }e^{-k x} \Psi ( a,1+c,-2 k x)\ ,\ \ \ \
\label{qafue}
\eeq
$$
b= \ha(-1+c)\ ,\ \  a=\ha (1+c)-{\a  \E \over 2k} \ ,\ \ c= \sqrt{ 4\mu^2-1-\a^2 }\ .
$$
The approximate energy eigenvalues in this point-like limit are  then determined by the formula
\beq
   -n_0= \ha (1+\sqrt{ 4\mu^2-1-\a^2 })-{\a  \E\over 2\sqrt{ {\cal M}^2-(\E-\ha\a )^2}}\ ,\ \ \ n_0=0,1,2,...
\label{atomo}
\eeq
Some eigenvalues are given in Table 3. Note that the binding energies are very small compared
to the asymptotic mass $\E_\infty={\cal M}+\ha \a $. This can be understood as follows. {}From eq.~(\ref{atomo}) we see that
\beq
\Delta \E =\E_\infty - \E \cong {\a^2(2{\cal M}+\a )^2\over 8{\cal M}(2n_0+1+c)^2} \ ,\ \ \
\E_\infty={\cal M}+\ha\a\ .
\eeq
This is of $O(\a^2) $ if ${\cal M}=O(1)$, and of   $O(\a^3) $ if ${\cal M}=O(\a  )$.
For example, for $\mu=1,n_0=0 $ and ${\cal M}=\a  $ one has  $ \Delta \E = 0.15 \ \a^3$.

 \begin{table}[htbp]
\centering
\begin{tabular}{l|l|ccc}
l & $n_0$ &
$k_{\rm num}$(s) & $k_{\rm num}$(a) & $k_{\rm point}$  \\
 \hline
1/2 & 0 & 0.03626  & 0.03728 & 0.03666  \\
1/2 & 1 &  0.02104 & 0.02140 &  0.02117    \\
3/2 & 0 &  0.02053 & 0.02053 & 0.02053   \\
3/2 & 1 &  0.014553 & 0.014555 & 0.014545    \\
5/2 & 0 & 0.0144460 & 0.01460 & 0.014460 \\
5/2 & 1 &  0.011217 & 0.011217 & 0.011217 \\
\end{tabular}
\parbox{5in}{\caption{Eigenvalues for binding energies   $k_{\rm num}$(s,a)
for $\eta_{\rm s}$ and $\eta_{\rm a}$ using $\a =0.2$ and ${\cal
M}=0.5$. $k_{\rm point}$ are the approximate eigenvalues given by
the analytic expression (\ref{atomo}). \label{tab:Efum}}}
\end{table}

\begin{figure}[hbt]
\label{fig3} \vskip -0.5cm \hskip -1cm
\centerline{\epsfig{figure=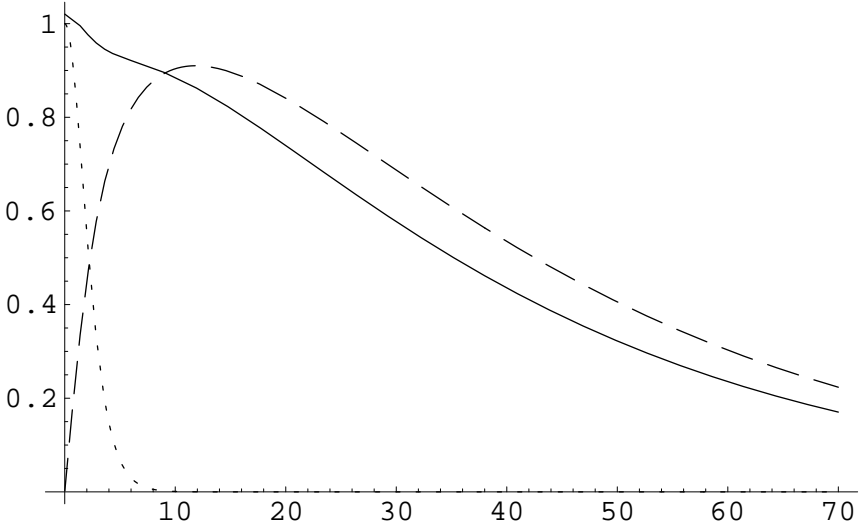,height=7truecm}}
\caption{\footnotesize Wave function for the lowest energy states
of $\eta_{\rm a}$ (solid line) and $\eta_{\rm s}$ (dashed line)
 $l=1$, $n_0=0$, for $\a =0.2 $ and ${\cal M}=0.5$.}
\end{figure}

\medskip

In the region $x\ll 1$, the equation becomes
\beq
-{1\over x^2}\p_x x^2\p_x
  \eta_{\rm s}   +\bigg({\mu (\mu+1)\over x^2}-k_{\rm in}^2\bigg)\eta_{\rm s}  = 0\ ,
\label{qqzzv}
\eeq
with
$$
\ k_{\rm in}^2=
-{\cal M}^2+ \E ^2+{ (1+\mu ) \over  6}\ .
$$
The solutions are Bessel functions.

 The numerical calculation of the eigenvalues is performed as in the previous cases by
integrating the differential equation from $x=0$ up to some $x=x_{\rm max}\gg 1 $,
with regular boundary conditions at zero. The  energy eigenvalues then
follow from the requirement that the wavefunction approaches zero at $x=x_{\rm max}$.
 Some results are given in Table 3.

\medskip

As in the case of the previous subsection, for the states of $\eta_{\rm a}$ with $\mu =1$ (i.e. $l=\ha $), the potential has no centrifugal barrier at
$x=0$, and the corresponding wave functions do not vanish at the origin.
The equation in the interior  $x\ll 1 $ becomes
\beq
   -{1\over x^2}\p_x x^2\p_x
  \eta _{\rm s} - k_{\rm in}^2\eta_{\rm s}  = 0\ ,
\label{qzzrt}
\eeq
with
$$
\ k_{\rm in}^2=-{\cal M}^2+ \E ^2\ .
$$
Although there is no centrifugal barrier near $x=0$, it is worth noting that
the particle is not concentrated at the core.  Its extension grows like
$$
\Delta x \cong {1\over\Delta \E} \cong {1\over {\cal M}\a^2 }\ ,
$$
which is much greater than one for small coupling $\a $.

The wave functions for the lowest energy states of $\eta _{\rm s}$ and
$\eta _{\rm a}$ are shown in fig.~3, for the case $\a =0.2$, ${\cal M}=0.5$.
Comparing to the case of fig.~2, where the Coulomb interaction is dominated by the coupling to the Higgs field, we see that the wave function is more extended. This is expected, in view of the above observation
that the extension is greater for smaller couplings.

The lowest energy state for $\eta_{\rm s}$ should be unstable
because the $\eta_{\rm s}$ entangled state can make a transition to a $\eta_{\rm a}$ entangled state by emission of a photon (see also  the discussion in Section 3).
However, the lowest energy bound state for $\eta_{\rm a}$ (with $l=\ha $, $n_0=0$) should be stable, since there is no possibility
of decay into a state of lower energy.
The same conclusion applies for the bound states of Section
4.3 .

\section{Summary and Discussion}


The main problem to which we addressed our study was the nature of the
possible bound states of a scalar particle around a dyon,
in systems which are globally electrically neutral.
Formally, this amounts to considering the scalar field
(different from the Higgs) as a fluctuation in the background of the dyon
solution, and treating it in the second quantization formalism.
Since we have taken as a paradigm of the dyon the  Prasad-Sommerfeld
solution of $N=2$ supersymmetric $SU(2)$ YM theory, we first
considered  scalar fields in $N=2$ supermultiplets. In these cases,
the scalar particle has an attractive interaction with the
long range Higgs field, whose strength is fixed by the dyon solution,
besides the usual Coulomb force (which is proportional to the square of the electric charge,
and  we naturally take it to be quite weaker than the other).
In Section 4.4, we considered a case where the scalar field
does not interact with the Higgs, which can be the case
in systems with less supersymmetry. Here the bound state is solely due to the Coulomb attraction. The main difference with respect to the previous cases is that bound states are, in this case, much more extended far from the monopole core. This is  expected, since the Coulomb interaction is weaker.

It should be noted that the bound states are rather larger than the
monopole core even for the cases where the Higgs attraction is dominant. Therefore they are expected
to be robust against  back-reaction effects of the classical
dyon solution for sufficiently small coupling constant
(apart from the effect of charge conservation which we have
already included).

In the case of the scalar field in the adjoint representation,
we have considered it to be a member of the $N=2$ vector multiplet.
Therefore it is coupled to the Higgs and it takes mass through the
Higgs expectation value. The attractive force is dominated by
the Higgs field
and  there are bound states also in the limit of vanishing charge.
The bound states correspond to a
certain linear combination $|{\rm dyon}_+\rangle |{\rm scalar}_-\rangle
- |{\rm dyon}_-\rangle |{\rm scalar}_+\rangle $, whereas the other
linear combination $|{\rm dyon}_+\rangle |{\rm scalar}_-\rangle
+ |{\rm dyon}_-\rangle |{\rm scalar}_+\rangle $ mixes with the state
$|{\rm monopole}_0\rangle |{\rm scalar}_0\rangle$,
in which both the particle and the monopole are uncharged.
This mixing cannot form bound states because
the neutral particle
is massless and
therefore it can have an arbitrarily low energy and still
escape to infinity. Since some perturbation not included in
our analysis, say the e.m. quanta radiation, could
cause a transition from
$|{\rm dyon}_+\rangle |{\rm scalar}_-\rangle
- |{\rm dyon}_-\rangle |{\rm scalar}_+\rangle $ to $|{\rm dyon}_+\rangle |{\rm scalar}_-\rangle
+ |{\rm dyon}_-\rangle |{\rm scalar}_+\rangle $,
we conclude that these bound states are  unstable.

In the case of the scalar field in the fundamental representation,
we have considered two cases. In the first case, the scalar is in a  $N=2$ hypermultiplet and is coupled
to the Higgs field by maintaining the $N=2$ supersymmetry of the Lagrangian.
We have assumed that its mass is completely due to
the Higgs expectation value, in order to compare with the previous case.
The resulting bound states are again essentially due to the
Higgs attraction.
In the second case, we have explicitly studied a SUSY breaking scenario,
in which the coupling to Higgs is  absent and there is an arbitrary  mass parameter.
The qualitative features of the two cases are rather similar.

Since in the fundamental representation there are no
particles of zero charge, both  combinations
$|{\rm dyon}_+\rangle |{\rm scalar}_-\rangle
\pm |{\rm dyon}_-\rangle |{\rm scalar}_+\rangle $
admit bound states (here we consider dyons with the same $\pm e/2$ charges
as the particles in the fundamental representation, in order to
have a globally neutral system). The lowest energy level is of
the form $|{\rm dyon}_+\rangle |{\rm scalar}_-\rangle
- |{\rm dyon}_-\rangle |{\rm scalar}_+\rangle $, with
the angular momentum taking the minimum value $l=\ha $.
In this case the ``interior centrifugal barrier'' is weak
and the wave function is nonzero at the origin, where typically
it takes its maximum value
(an exception occurs in the second case with no coupling to Higgs, when the charge is very small: in such case, we found that the maximum of the wave function appears at some finite $r$).
The extension of the wave function is in both cases larger
than the monopole core size.
The orbiting particle can thus be always considered to be mostly outside the core, with a wave function similar to that of the Hydrogen atom. The main role of the monopole core is to produce
a full quantum entanglement, despite its small effect on the $r$ dependence of the wave function.

Thus a bona-fide bound state picture  emerges in both cases. We
conclude that we get  stable, quantum entangled, bound states in
the case of the fundamental representation.

While generic quantum entanglement of charges may occur in various
physical situations (for instance in the case of the decay of a
neutral particle in two charged components), the remarkable
phenomenon found here is that stable bound states of a light
particle and a heavy dyon can only exist in an entangled state
involving  the system with opposite charge. The bound particle and
the dyon can be very far from each other, and they are not charge
eigenstates (although the total charge is zero). Although this
setting does not seem to be in conflict with any physical law, it
is nevertheless a curious effect and one could amuse
her-(him-)self by imagining would be paradoxes in the chemistry of
such an ``atom''.

Note that a measurement of the charge of the orbiting particle should make the wavefunction precipitate into a defined charge eigenstate
(say, from $|{\rm dyon}_+\rangle |{\rm scalar}_-\rangle
- |{\rm dyon}_-\rangle |{\rm scalar}_+\rangle $ to
$|{\rm dyon}_+\rangle |{\rm scalar}_-\rangle $).
Since such bound state is not possible, one concludes that any measurement of the charge should require an energy above the binding energy, so that the final scalar particle can escape to infinity.

Similar bound states should exist for
fermions and vector particles
 i.e. they should give rise to a state
consisting of a mixture of fermions or
 $W^+$ and $W^-$ gauge bosons entangled with dyons.
  This may be more easily derived by embedding the $N=2$ supersymmetric model
in $N=4$ super Yang-Mills theory,
so that the solutions of the wave equation for fermion and gauge boson fluctuations  are connected by an unbroken
supersymmetry to the solutions for  scalar fields computed here.

\section{Acknowledgements}

J.R. wishes to thank SISSA and ICTP for hospitality and for a financial support.
R.I. acknowledges partial support from the EEC contract HPRN-CT-2000-00131.

\section{Appendix: Harmonic Analysis for adjoint scalars}

In this Appendix we perform the harmonic analysis of the coupled equations
for the case of scalar particles in the adjoint representation (the harmonic analysis
in the case of a pointlike monopole has been done in ref.\cite{yang}).

Using eqs.~(\ref{udieci}) - (\ref{ffpm}),
we obtain  the following system of equations in spherical coordinates:
\beqa
 && \vec   D^2 \cdot \big( \hat\g \ F_0(\vec r) \big) =  \hat \g  \ \bigg(
 {1\over r^2}\p_r r^2\p_r
-{2K^2\over r^2}-{1\over r^2}\hat L^2\bigg) F_0
\non\\
&-& \hat\a_- \  {K \over r^2\sin^2\theta } (\sin\theta \p_\theta -i\p_\vp )  F_0 -
 \hat \a_+ \   {K \over r^2 \sin^2\theta } (\sin\theta \p_\theta + i\p_\vp )  F_0 \ ,
\label{uci}
\eeqa
\beqa
\vec   D^2 \cdot \big[ \hat\a_ \mp F_\pm \big] &=&
\hat\a _\mp  {e^{\pm i\vp}\over\sin\theta }\big[ {1\over r^2}\p_r r^2\p_r
 -{1\over r^2}(-1+K^2)-{1\over r^2}\hat J^2_{\pm} \big]h_\pm
\non\\
&+&\hat\g\ {2K\over r^2 } {e^{\pm i \vp}\over \sin\theta }\bar \nabla_\pm h_\pm\ ,
\label{quf}
\eeqa
where
\beq
\hat L^2=-{1\over \sin^2\theta }(\sin\theta\p_\theta+i\p_\vp )
(\sin\theta\p_\theta-i\p_\vp )\ ,
 \label{L2}
\eeq is the standard angular momentum and \beq \hat
J^2_{+}=-{1\over \sin^2\theta } \nabla _+ \bar \nabla _+\ , \ \ \
\  \hat J^2_{-}=-{1\over \sin^2\theta } \nabla _- \bar \nabla _-
\ , \label{J2} \eeq is a covariant angular momentum. Here
$$
\nabla_\pm =\sin\theta \p_\theta  \mp i\p_\vp +(1-\cos\theta )\ ,
$$
$$
\bar \nabla_\pm =\sin\theta \p_\theta  \pm i\p_\vp -(1-\cos\theta )\ ,
$$
$$
[\nabla_\pm,\bar \nabla_\pm ]=-2\sin^2\theta\ .
$$

The full equations of motion can now be rewritten as
\beq
 \bigg( -{1\over r^2}\p_r r^2\p_r
+{2K^2\over r^2}+{1\over r^2}\hat L^2\bigg) F_0
- {2K\over r^2 \sin\theta  }\big[ e^{i \vp}\bar \nabla_+ h_+   +
 e^{-i \vp}\bar \nabla_- h_-\big]
 =E^2 F_0     \ ,
\label{zaaii}
\eeq
\beqa
  & \big[ -{1\over r^2}\p_r r^2\p_r
 +{1\over r^2}(-1+K^2)+{1\over r^2}\hat J^2 \big]h_+
 + {K e^{-i\vp }\over r^2\sin\theta } (\sin\theta \p_\theta -i\p_\vp )  F_0
\non\\
& =\big[ (E  -e V \hat q )^2 - e^2 a^2  G^2\big] h_+\ ,
\label{zaapi}
\eeqa
 \beqa
  & \big[ -{1\over r^2}\p_r r^2\p_r
 +{1\over r^2}(-1+K^2)+{1\over r^2}\hat J^2 \big]h_-
 + {K e^{i\vp }\over r^2\sin\theta } (\sin\theta \p_\theta +i\p_\vp )  F_0
\non \\
& = \big[ (E  + e V \hat q  )^2 - e^2 a^2  G^2\big] h_-\ .
\label{zaapp}
\eeqa

The angular dependence is solved by setting
\beq
F_0(\vec r)= \phi_0(r)\ Y_{lm}(\theta,\vp )\ ,\ \
\eeq
\beq
h_+ (\vec r)= {1\over l_0}\ \phi_+ (r)\ Z_{lm}^+(\theta,\vp )\ ,\ \ \ \
h_- (\vec r)={1\over l_0 }\ \phi_- (r)\ Z_{lm}^-(\theta,\vp )\ ,\
\label{uia}
\eeq
\beq
\ l_0\equiv\sqrt{l(l+1)}\ ,
\eeq
where $ Y_{lm}(\theta,\vp )$ are the spherical harmonics, i.e.
\beq
\hat L^2 Y_{lm}=l(l+1) Y_{lm}\ ,
\eeq
with $l=0,1,2\dots$,
    and $Z_{lm}^\pm $ are eigenfunctions of the operator $\hat J^2 $,
\beq
\hat J^2 Z_{lm} = l(l +1) Z_{lm} \ .
\label{ssw}
\eeq
Explicitly, they are given by
\beq
Z_{lm}^{\pm}= {e^{\mp i\vp }\over \sin\theta } (\sin\theta \p_\theta\mp i \p_\vp )Y_{lm}\ .
\label{dasa}
\eeq
Using (\ref{dasa}), one can check that  (\ref{ssw}) is indeed satisfied.

Here $l=1,2,\dots$.
Note that there is no eigenfunction of $\hat J^2$ for $l=0$, since there are no normalizable solutions
of  $\bar \nabla _{\pm} Z_0=0$.

Using eqs.~(\ref{zaaii})-(\ref{uia})
 and  the properties:
$$
{e^{i\vp }\over\sin\theta } \bar \nabla _+ Z_{lm}^+= -l(l+1) Y_{lm}\ ,\ \ \ \
{e^{-i\vp }\over\sin\theta } \bar \nabla _- Z_{lm}^-= -l(l+1) Y_{lm}\ ,
$$
we find eqs.~(\ref{aazi})-(\ref{aapp}).

The functions $Z_{lm}^{\pm}$ (\ref{dasa})
have a simple form in terms of $\theta ,\varphi $.
For example,
for $l=1$ we have the following solutions
$$
m=0:\ \ \ \ \ F_0(\vec r)=\phi_0(r)\cos\theta\ ,\ \ \
h_\pm(\vec r)={1\over\sqrt{2}} \ \phi _\pm(r) e^{\mp i\vp }\sin\theta
$$
$$
m=1:\ \ \ \ F_0(\vec r)=\phi_0(r)e^{i \vp }\sin\theta\ ,\ \ \ h_+(\vec r)={1\over\sqrt{2}}\phi_+(r) (1+\cos\theta ) \ ,\ \ \
$$
$$
 h_-(\vec r)= {1\over\sqrt{2}}  \phi_-(r) e^{2i\vp }(-1+\cos\theta ) \ ,\
$$
$$
m=-1: \ \ \ \ F_0(\vec r)=\phi_0(r)e^{-i \vp }\sin\theta\ ,\ \ \
h_+(\vec r)={1\over\sqrt{2}}\phi_+(r) e^{-2i\vp}(-1+\cos\theta ) \ ,\ \ \
$$
$$
 h_-(\vec r)={1\over\sqrt{2}}\phi_-(r) (1+\cos\theta ) \ .
$$
Similarly, one can write down expressions for higher $l$.

\end{document}